\documentclass[useAMS,usenatbib,usegraphicx]{mn2e}

\newcommand{\der}{{\rm d}} 
\newcommand{\aj}{AJ} \newcommand{\mnras}{MNRAS}
\newcommand{\apj}{ApJ} \newcommand{\apjl}{ApJ} 
\newcommand{\nat}{Nature} \newcommand{\apjs}{ApJS} 
\newcommand{\aap}{A\&A}  
\newcommand{\tir}{\tilde r}

\newcommand{\R}{{R_{\rm f}}}

\newcommand{\co}{_{\rm c}} \newcommand{\cs}{_{\rm pk}}
  
\newcommand{\rad}{_{\rm r}} 
 \newcommand{\DM}{_{\rm DM}}
 \newcommand{\HH}{^{\rm H}}

\newcommand{\pk}{_{\rm pk}} 
 \newcommand{\m}{_{-1}} 
\newcommand{\rs}{r_{\rm s}} \newcommand{\Ms}{M_{\rm s}}
\newcommand{\rsMsb}{$\Ms$--$\rs$\ } 
 \newcommand{\tot}{}

\newcommand{\p}{_{\rm p}} 

\newcommand{\modot}{M$_\odot$\ } \newcommand{\modotc}{M$_\odot$}
\newcommand{\ti}{t_{\rm i}} 
\newcommand{\ii}{_{\rm i}} \newcommand{\sphc}{(r,\theta,\varphi)}
 
 \newcommand{\nbody}{{$N$}-body }
\newcommand{\beq}{\begin{equation}} \newcommand{\eeq}{\end{equation}}
 \newcommand{\beqa}{\begin{eqnarray}}
\newcommand{\eeqa}{\end{eqnarray}} \newcommand{\lav}{\langle}
\newcommand{\rav}{\rangle} \newcommand{\col}{_{\rm ta}}
\newcommand{\vir}{_{\rm vir}} \newcommand{\spot}{\lav\Phi\rav}
\newcommand{\srho}{\lav\rho\rav}

\newcommand{\isph}[1]{\int_0^{2\pi}\!\!\der\varphi\!\!
\int_0^{\pi}\!\!\der\theta\,\sin\theta\;{#1}}
\newcommand{\derpr}{\partial_{\rm r}}

\begin{document}

\title[Halo Density Profile] {Theoretical dark matter
  halo density profile}

\author[Salvador-Sol\'e at al.]{Eduard Salvador-Sol\'e\thanks{E-mail: e.salvador@ub.edu}, Jordi Vi\~ nas, Alberto Manrique and Sinue Serra
\\Institut de Ci\`encies del Cosmos,
Universitat de Barcelona (UB--IEEC), 
Mart{\'\i} i Franqu\`es 1, E-08028 Barcelona, Spain}


\maketitle
\begin{abstract}
We derive the density profile for collisionless dissipationless dark
matter haloes in hierarchical cosmologies making use of the Secondary
Infall (SI) model. The novelties are: i) we deal with triaxial
virialised objects; ii) their seeds in the linear regime are peaks
endowed with {\it unconvolved} spherically averaged density profiles
according to the peak formalism; iii) the initial peculiar velocities
are taken into account; and iv) accreting haloes are assumed to
develop from the inside out, keeping the instantaneous inner system
unaltered. The validity of this latter assumption is accurately
checked by comparing analytical predictions on such a growth with the
results of numerical simulation. We show that the spherically averaged
density profile of virialised objects can be inferred with no need to
specify their shape. The {\it typical} spherically averaged halo
density profile is inferred, down to arbitrarily small radii, from the
power-spectrum of density perturbations. The predicted profile in the
$\Lambda$CDM cosmology is approximately described by an Einasto
profile, meaning that it does not have a cusp but rather a core, where
the inner slope slowly converges to zero. Down to one hundredth the
total radius, the profile has the right NFW and Einasto forms, being
close to the latter down to a radius of about four orders of magnitude
less. The inner consistency of the model implies that the density
profiles of haloes harbour no information on their past aggregation
history. This would explain why major mergers do not alter the typical
density profile of virialised objects formed by SI and do not
invalidate the peak formalism based on such a formation.
\end{abstract}



\section{INTRODUCTION}\label{intro}

In the last two decades, observations and, particularly simulations,
have provided us with detailed information on the structure and
kinematics of bottom-up hierarchically assembled virialised dark
matter haloes. However, from the theoretical viewpoint, the situation
is far from satisfactory. The way all these properties settle down
remains to be elucidated and their connection with the power-spectrum
of density perturbations is unknown.

The accurate modelling of virialised self-gravitating collisionless
dissipationless systems is an old unresolved issue. Most efforts
have focused on determining the equilibrium density profile for
objects formed from the monolithic collapse of an isolated spherically
symmetric seed with outward-decreasing density profile and pure Hubble
velocity field, the so-called Secondary Infall (SI) model. Following
the seminal work by \citet{GG72}, the power-law density profile was
derived under the self-similar approximation and/or making use of an
adiabatic invariant during virialisation, both for pure radial orbits
\citep{G77,FG84,Ber85,HS85} and non-radial orbits
\citep{RG87,WZ92,Nuss01,LDH03}. The departures from spherical symmetry
(by adopting, instead, cylindrical symmetry; \citealt{Ry93,Mout95})
and self-similarity \citep{Lok00,LH00} were also investigated.

These theoretical results were tested and complemented by the
information drawn from specifically designed numerical tools
\citet{G75,Wea04} as well as full cosmological \nbody simulations
(e.g. \citealt{Fea85,QSZ86,Eea88,Zea96}). One important finding along
this latter line was that cold dark matter (CDM) haloes with different
masses show similar scaled spherically averaged density profiles
\citep{dc91,Cea94}. \citet{NFW97} showed that they are well-fitted
down to about one hundredth of the total radius by a simple analytic
expression, the so-called NFW profile, that deviates from a
power-law. This finding opened a lively debate about the value of the
central asymptotic behaviour of the halo density profile
(e.g. \citealt{FM97,Moore98,Gea00,JS00,P03,Hayaea04}).  More recently,
the \citet{E65} profile was shown to give even better fits down to
smaller radii \citep{Navea04,Navea10,M05,M06}.

The origin of this density profile remains to be
understood. Certainly, it can only arise from the way haloes aggregate
their mass, which, in hierarchical cosmologies, is through continuous
mergers. But the dynamical effect of mergers depends on the relative
mass of the captured and capturing haloes, $\Delta M/M$. For this
reason, it is usually distinguished between minor and major mergers
($\Delta M/M$ greater or less than about 0.3; \citealt{smgh07}). Major
mergers have a dramatic effect each on the structure of the
object. While minor mergers contribute jointly, together with the
capture of diffuse matter (if any), to the so-called accretion that
yields only a small secular effect on the accreting object.  Some
authors \citep{SW98,RGS98,Suea00,Dea03} studied the possibility that
the non-power-law density profile for simulated haloes is the result
of repeated major mergers. Others
\citep{ARea98,Hea99,DPea00,As04,McM06} concentrated on the effects of
pure accretion (PA), that proceeds according to the simple SI model
above from peaks (secondary maxima) in the primordial random Gaussian
density field \citep{D70,BBKS}. The density profile found in this
latter scenario appears to be similar, indeed, to that of simulated
haloes. Furthermore, recent cosmological simulations have confirmed
that major mergers play no central role in setting the structure and
kinematics of dark matter haloes \citep{WW09}. However, the reason why
major mergers would not alter the density profile set by PA is not
understood.

In the present paper, we develop an accurate model of halo density
profile that clarifies all these issues. For simplicity, we consider
pure dark matter systems, that is we neglect the effects of baryons on
halo structure. This will allow us to directly compare the theoretical
predictions of the model to the results of $N$-body simulations.  This
model is built within the SI framework and uses the peak formalism,
which assumes there is a one-to-one mapping between haloes and the
density peaks. However, peaks are not assumed to be spherically
symmetric and endowed with their typical {\it filtered} density
profile calculated by \citeauthor{BBKS} (\citeyear{BBKS}, hereafter
the BBKS profile) as usual, but we take into account that they are
triaxial \citep{D70,BBKS} and consider their accurate spherically
averaged {\it unconvolved} density profile. In addition, we account
for the initial peculiar velocities believed to affect the central
halo density profile \citep{H02,As04,MDL07}. Finally, instead of
making use of the usual adiabatic invariant, poorly justified near
turnaround and greatly complicating the analytic derivation of the
final density profile, we take advantage of the fact that accreting
haloes develop outwardly, keeping their instantaneous inner structure
unaltered. The validity of such an assumption, hereafter simply
referred to as inside-out growth, is accurately checked by verifying
that the trends observed in simulated haloes evolving by smooth
accretion are reproduced by an analytic model where the haloes are
forced to grow inside-out.

And what about the fact that major mergers are ignored in SI? This
important caveat affects not only the modelling through SI of the halo
density profile, as mentioned, but also of halo statistics in the
so-called peak formalism (\citealt{PH90,bm,MSS95,MSS96}, hereafter
MSSa and MSSb, respectively; \citealt{metal98}) which relies on the
Ansatz that there is a one-to-one correspondence between haloes and
peaks as in spherical collapse\footnote{The excursion set formalism is
  also based on a one-to-one correspondence between haloes and
  overdense regions in the initial density field as in spherical
  collapse.}. Although no such one-to-one correspondence is actually
found in simulations \citep{Kea93}, this is because peaks show nested
configurations (MSSa) which are not corrected for. If one concentrates
in simply checking whether or not halo seeds, hereafter also called
protohaloes, coincide with peaks, then the answer is positive in all
but a few percent of cases compatible with the frequency of non-fully
virialised haloes \citep{Pea}.  However, there is still the problem
that, in principle, major mergers should blur the correspondence
between peaks and haloes arising from SI.  In the present paper, we
argue that the reason why halo statistics and the spherically average
density profile of virialised haloes seem to be unaffected by major
mergers is that virialisation is a real relaxation process. We stress
that the model deals with {\it fully} virialised haloes. This is
important because, after a major merger, the density profile of a halo
spends some time to adopt a stable density profile.

The outline of the paper is as follows. In Section \ref{general}, we
present some general relations holding for all systems regardless of
their symmetry. Taking into account these relations, we develop in
Section \ref{model} the model for triaxial haloes grown by PA. In
Section \ref{peaks}, we calculate the properties of typical
protohaloes and use them in Section \ref{haloes} to derive the typical
spherically averaged density profile for CDM haloes. In Section
\ref{discussion}, we discuss the effects of major mergers. The main
results are summarised in Section \ref{summ}.

Throughout the paper, we adopt the $\Lambda$CDM Wmap7 \citep{Km11} 
concordance model. In spite of this, we use Newtonian
dynamics with null cosmological constant as its effects are irrelevant
at the scale of virialised haloes. A package with the numerical codes
used in the present paper is publicly available from {\texttt
  {www.am.ub.es/$\sim$cosmo/haloes\&peaks.tgz}}.

\section{General Relations for Spherical Averaged Profiles}\label{general}

In the present section, we derive some general relations for
spherically averaged quantities that hold for any arbitrary system
regardless of its symmetry and that will later be used to build the
model.

Consider a self-gravitating system with arbitrary mass distribution,
aggregation history and dynamical state. The local density and
gravitational potential can be split in the respective spherical
averages around any given point and the corresponding residuals,
\beqa
\,\rho\sphc=\srho(r)+\delta\rho\sphc
\label{split1}~~~\\
\Phi\sphc =\spot(r)+\delta\Phi\sphc\,.
\label{split2}
\eeqa
The spherically averaged gravitational potential,
\beq
\spot(r) = \frac{1}{4\pi}\!\isph{\Phi\sphc} ,
\label{spot}
\eeq
satisfies, by the Gauss theorem, the usual Poisson integral relation
for spherically symmetric systems,
\beq
\frac{\der\spot(r)}{\der r} = \frac{G M(r)}{r^2}\,,
\label{gauss}
\eeq
where $M(r)$ is the mass within $r$, 
\beq
M(R)=4\pi\int_0^{R} \der r\, r^2\,\srho(r)\,.
\label{mass}
\eeq

Taking into account the null spherical averages of $\delta\rho$ and
$\delta\Phi$ (see eqs.~[\ref{split1}]--[\ref{split2}]), the potential
energy within the sphere of radius $R$,
\beq 
W(R)=\frac{1}{2}\!\int_0^R \!\der r\,r^2\!\!\isph{\!\rho\sphc\,\Phi\sphc}\,,
\label{potential}
\eeq
can be rewritten as
\beqa
W(R)=2\pi\!\int_0^R\!\! \der r\,r^2[\srho(r)\,\spot(r) + \lav\delta\rho\,\delta\Phi\rav]\nonumber~~~~~~~~~~~~~\\
    =-4\pi\!\!\int_0^R\!\! \der r\,r^2\srho(r)\,\frac{GM(r)}{r} + 2\pi\!\!\int_0^R \!\!\der r\,r^2\lav\delta\rho\,\delta\Phi\rav\nonumber\\
\equiv {\cal W}(R) + \delta {\cal W}(R)\,,~~~~~~~~~~~~~~~~~~~~~~~~~~~~~~~~~~~~~~~
\label{potexac}
\eeqa
where we have introduced the ``spherical'' potential energy within
$R$,
\beq
{\cal W}(R)=-4\pi\int_0^R \der r\,r^2\srho(r)\,\frac{GM(r)}{r}\,.
\label{spot2}
\eeq
The second equality in equation (\ref{potexac}) follows from
partial integration of the first term on the right of the first
equality, then application of the relation (\ref{gauss}), and one new
partial integration, choosing the origin of the spherically averaged
potential so as to have
\beq
\spot(R)=-\frac{GM(R)}{R}\,.
\label{opot}
\eeq 
Note that, for spherically symmetric systems, this boundary condition
coincides with considering the usual potential origin at infinity and
the system truncated at $R$.

The kinetic energy within $R$ is
\beq
K(R)=2\pi\int_0^R \der r\,r^2\srho(r)\, \sigma^2(r)\,,
\label{kinexac}
\eeq
being $\sigma^2(r)$ the velocity variance in the infinitesimal shell
at $r$.\footnote{Such a velocity variance coincides with the spherical
  average of the local value. Therefore, equation (\ref{kinexac})
  could also be written in terms of the local velocity variance in
  angular brackets.} Thus, the total energy of the sphere,
$E(R)=K(R)+W(R)$, takes the form
\beqa
E(R)\!=4\pi\! \int_0^R \der r\,r^2\srho(r) \left[\frac{\sigma^2(r)}{2}-\frac{GM(r)}{r}\right]\nonumber~~~\\+\, 
2\pi\!\int_0^R \der r\,r^2\lav\delta\rho\,\delta\Phi\rav(r)
\equiv {\cal E}(R) + \delta {\cal E}(R)
\,,
\label{ener}
\eeqa
where we have introduced the ``spherical'' total energy
\beq 
{\cal E}(R)= 4\pi \int_0^{R} \der
r\,r^2\srho(r)\left[\frac{s^2(r)}{2}-\frac{GM(r)}{r}\right]\,.
\label{energy}
\eeq
Note that ${\cal E}(R)$ is written in terms of the ``spherical''
velocity variance $s^2(r)$, different, in general, from the ordinary
velocity variance,
\beq
\sigma^2(r)=s^2(r)+\delta s^2(r)\,,
\label{residual}
\eeq
through the residual $\delta s^2(r)$ to be specified, contributing to
the residual $\delta {\cal E}(r)$. In other words, there is some
freedom in the definition of $s^2(r)$. We will come back
to this point later.

So far we have considered a system with arbitrary mass distribution
(symmetry), aggregation history and dynamical state. If the system is
in addition in equilibrium, then multiplying the steady collisionless
Boltzmann equation (e.g. eq.~[4p-2] in \citealt{bt87}) by the radial
particle velocity and integrating over velocity and solid angle, we
are led to
\beqa
\frac{\der (\srho \sigma\rad^2)}{\der r}+ \frac{\srho(r)}{r} 
[3\sigma\rad^2(r) -
\sigma^2(r)]\nonumber~~~~~~~~~~~~~~~~~~~~~~~ \\= -\frac{1}{4\pi}\isph{\rho\sphc\,\derpr\Phi\sphc}\,,
\label{Jeq1}
\eeqa
where $\derpr$ stands for radial partial derivative\footnote{To
derive equation (\ref{Jeq1}) it is only needed such conventional
conditions as the continuity of the local density and mean velocity
and the fact that the velocity distribution function vanishes for
large velocities.}. Taking into account equation (\ref{gauss}),
equation (\ref{Jeq1}) adopts the form
\beqa 
\frac{\der (\srho
\sigma\rad^2)}{\der r}+ \frac{\srho(r)}{r} [3\sigma\rad^2(r) -
\sigma^2(r)]\nonumber~~~~~~~~~\\ = -\srho(r)\frac{GM(r)}{r^2} -
\lav\delta\rho\,\derpr(\delta\Phi)\rav(r)\,,
\label{exJeq}
\eeqa
identical to the Jeans equation for spherically symmetric systems in
equilibrium but for the last term on the right. Multiplying equation
(\ref{exJeq}) by $4\pi r^3$ and integrating over the sphere of radius
$R$, the same steps leading to the scalar virial relation for
spherically symmetric self-gravitating systems now lead to
\beqa
4\pi R^3\srho (R)\sigma\rad^2(R) - 2K\nonumber~~~~~~~~~~~~~~~~~~~~~~~~~~~~~~~~~~~~~~~~\\ =\!-4\pi\!\!\int_0^R\!\der r \,r^2\srho(r)\frac{GM(r)}{r}\! -\! 4\pi\!\! \int_0^R\!
\der r\, r^3 \lav\delta\rho\, \derpr \delta\Phi\rav(r) \,.
\label{vir1}
\eeqa
Therefore, defining the so-called ``spherical'' radial velocity
variance $s^2\rad(r)$ through
\beq
\sigma\rad^2(r)=s^2\rad(r)+\delta s^2\rad(r)\,,
\label{radial}
\eeq
with
\beq 
\delta s^2\rad(r)= \frac{1}{r^3\srho(r)}\int_0^r \der \tir\,
\tir^2 \left[\delta s^2(\tir)-\tir \lav\delta\rho\, \derpr \delta\Phi\rav (\tir)\right]\,,
\label{dsr}
\eeq
the virial relation (\ref{vir1}) takes the usual form for spherically
symmetric systems,
\beq
4\pi R^3\srho(R)\, s^2\rad(R)=4\pi\!\!\int_0^{R}\!\der
r\, r^2\srho(r)\!\left[s^2(r)\!-\!\frac{GM(r)}{r}\right]\!\!,
\label{vir2l}
\eeq 
from now on called the ``spherical'' virial relation. Furthermore,
defining the ``spherical'' scaled surface pressure term (from now on
simply spherical surface term), ${\cal S}(R)$, equal to the member on
the left of equation (\ref{vir2l}) over the absolute value of ${\cal
  W(R)}$, the spherical virial relation (\ref{vir2l}) adopts the usual
compact form
\beq 
\frac{2{\cal E}(R)}{{\cal W}(R)}=1-{\cal S}(R)\,.
\label{vir0l}
\eeq

As can be seen, all the ordinary quantities $X$ can be expressed in
the form $X={\cal X}+\delta {\cal X}$ with the quantities ${\cal X}$
having the same form for spherically symmetric systems, hence why they
are labelled ``spherical''. The residuals $\delta{\cal X}$ always
measure the deviation from sphericity of the quantities $X$ in the
sense that, when $\delta\rho$ and $\delta\Phi$ vanish, $\delta{\cal
  X}$ also vanish and $X$ become equal to the spherical counterparts
${\cal X}$ and, hence, recover the form for spherically symmetric
systems. The dimensionless quantities $|\delta {\cal X}/ {\cal X}|$
are, however, not necessarily less than one, so we should not look at
the spherical quantities ${\cal X}$ as 0th-order approximations of $X$
in expansion series on small deviations from sphericity; they are
fully exact quantities and so are also the relations between them.  In
particular, equations (\ref{mass}), (\ref{energy}) and (\ref{vir0l})
respectively giving the mass, spherical total energy and spherical
virial relation or, equivalently, the relations
\beq
\srho(r)= \frac{1}{4\pi r^2}\frac{\der M}{\der r}\,,
\label{rhot}
\eeq
\beq 
s^2(r)=2\left[\frac{\der {\cal E}/\der
r} {\der M/\der r}+ \frac{GM(r)}{r}\right],
\label{sig2}
\eeq
and
\beq 
s\rad^2(r)=\frac{2{\cal E}(r)-{\cal W}(r)}{r\,\der M/\der r}\,,
\label{sigr2}
\eeq 
following from their differentiation are exact and hold unchanged
regardless of the particular shape of the object. Thus, the profiles
$\srho(r)$, $s^2(r)$ and $s^2\rad(r)$ {\it do not depend on shape} and
can be determined by solving equations (\ref{rhot})--(\ref{sig2}). In
contrast, the ordinary kinematic profiles $\sigma^2(r)$ and
$\sigma^2\rad(r)$ cannot be obtained from a similar set of equations
because they depend explicitly on the shape of the object.

 We want to remark that, even if the spherical quantities {\it at any
   given moment} behave as their ordinary counterparts in spherically
 symmetric systems, the relation between their values {\it in
   different epochs} may not be the same as in such systems. For
 instance, the spherical total energy inside spheres of fixed mass may
 not be conserved in the absence of shell-crossing\footnote{If one
   chooses $s^2(r)\equiv \sigma^2(r)$, the spherical total energy is
   not conserved in ellipsoidal systems because any given sphere
   exchanges energy with the rest of the system.}. Only one specific
 choice of $s^2(r)$ will ensure such a conservation. Note also that,
 for any arbitrary mass distribution, it might not be possible to have
 one only profile $s^2(r)$ satisfying that condition for the whole
 series of embedded spheres the system decomposes into. But, in the
 case of a centred triaxial system, this is always possible thanks to
 the ordered radial mapping of the mass distribution. From now on, we
 assume such a symmetry and the conservation of the spherical
   total energy in the absence of shell-crossing. This is crucial for
 the model as it will allow us to fix the density profile for a
 virialised object from the properties of its seed (see
 Sec.~\ref{model}).

\section{The Model}\label{model}

We will concentrate, in a first step, on triaxial systems {\it evolving
  by PA}. This includes the case of accretion along filaments.

\subsection{Inside-out Growth}\label{inout}

The model relies on the assumption that accreting objects grow
inside-out. This is consistent with the properties of simulated haloes
\citep{ssm,SMS05,Hea03,RD06,Wangea11}. In \citet{metal03} and
\citet{smgh07}, it was shown that the assumption that (spherical)
haloes grow inside-out at the typical cosmological accretion rate
automatically leads to a density profile \`a la NFW. Nonetheless, to
be fully confident about it, we show below that such a growth is
naturally expected and in agreement with the detailed results of
simulations.

In SI, the isodensity contours in the triaxial seed expand, without
crossing each other, until they reach turnaround. This expansion is
achieved in linear regime with all particles moving radially, so the
axial ratios of shells are conserved. However, after reaching
turnaround, particles collapse and rebound non-radially, so the shape
of the isodensity contours changes. Shell-crossing causes a {\it
  secular} energy transfer between shells so that particles initially
lying in a turnaround ellipsoid reach their new apocentre at a smaller
radius and not simultaneously. However, the energy lost by particles in one
orbit is small, so the typical time of apocentre variation is
substantially greater than the typical orbital period, implying that
particles belonging to one turnaround ellipsoid will still trace an
effective apocentre surface, which for symmetry reasons will also be
an ellipsoid although smaller in size and with different axial
ratios. Repeating the same reasoning from each new apocentre
ellipsoid, we have that any turnaround ellipsoid evolves through a
continuous series of ellipsoidal apocentre surfaces that progressively
shrink and change their shape until they stabilise.

During such an evolution, apocentre ellipsoids never cross each
other. If two of them had a common point, particles belonging to the
two surfaces at that point (with null velocities) would follow the
same evolving orbits, so the two surfaces would always be in
contact, but this is meaningless because different turnaround
ellipsoids come from different isodensity contours in the seed which do
not intersect. Thus, during virialisation, the system contracts
orderly, {\it without apocentre-crossing} and when a new apocentre
ellipsoid stops shrinking, it necessarily places itself beyond all
previously stabilised apocentre ellipsoids. Thus, the central steady
object necessarily grows from the inside out according to the gradual
deposition of particles with increasingly larger apocentre surfaces.

In the above reasoning, we assumed that orbits eventually
stabilise. However, new shells are constantly arriving and crossing
the growing halo, consequently, orbits only approximately
stabilise. It is therefore worthwhile verifying the validity of this
approximation. This can be accurately checked by means of the detailed
follow-up of the density profile for CDM haloes in simulations.
\citet{ZJMB09} and \citet{Mea} carried out such a follow-up and found
that the $\rs$ and $\Ms$ NFW shape parameters\footnote{$\Ms$ is
  defined as the mass within the scale radius $\rs$ where the
  logarithmic slope of the density profile is equal to $-2$.} of
accreting haloes are not constant but vary slightly with time.  This
was interpreted as evidence that the inner structure of accreting
haloes is changing. But this is not the only possible
interpretation. As the density profile of haloes {\it is not perfectly
  fitted} by the NFW profile\footnote{The larger the mass, the closer
  the density profile is to a power-law with index $\sim -1.5$-$2$
  \citep{TKGK04}.}, even if the inner density profile did not change,
its fit by the NFW law out to progressively larger radii should result
in slightly different best values of $\rs$ and $\Ms$.

\begin{figure}
 \includegraphics[scale=0.45]{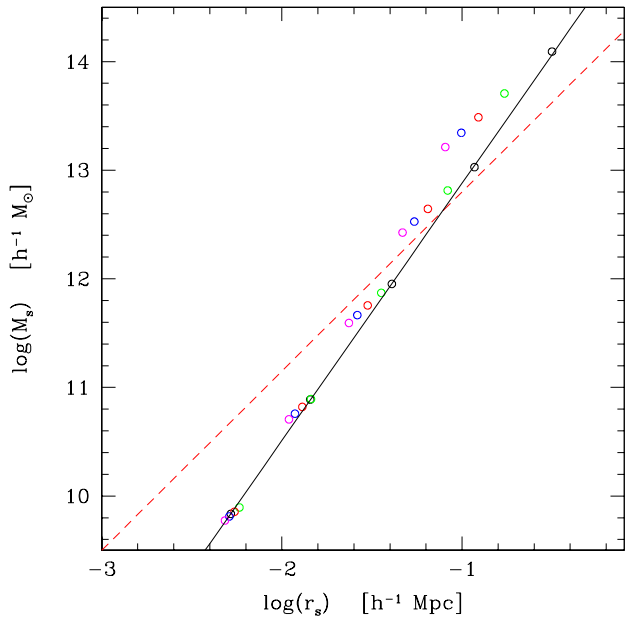}
 \caption{Tracks followed in the \rsMsb plane by typical accreting
   haloes developing inside-out, with current masses, $M\tot$, equal
   to $10^{11}$, $10^{12}$, $10^{13}$, $10^{14}$ and $10^{15}$ \modot
   (from bottom to top) at $z=0$, 2, 4, 6 and 8 (circles
     essentially from right to left, respectively in black, green,
   red, blue and magenta). The straight dashed red line shows the
   universal direction followed by such tracks, identical to that
   found by \citet{ZJMB09} in full cosmological simulations. The black
   straight line shows the \rsMsb relation obtained at $z=0$ also
   identical to that found in \citeauthor{ZJMB09}
   al.~(\citeyear{ZJMB09}; see Fig.~22).}
\label{f0}
\end{figure}

To confirm that our interpretation for this change is correct we have
followed in the \rsMsb log-log plane the evolution of accreting halos
forced to grow {\it inside-out} at the typical accretion rate given by
the excursion set formalism according to \citet{smgh07} model. The
result, in the same cosmology as used in \citet{ZJMB09}, is shown in
Figure \ref{f0}. Even if, by construction, haloes grow inside-out
without changing their instantaneous inner structure, when their
density profile is fitted by a NFW profile, they are found to move in
the \rsMsb log-log plane along one fixed direction over a distance
that increases with current halo mass. This behaviour is {\it
  identical} to that found by \citeauthor{ZJMB09}: accreting halos
moved in the \rsMsb log-log plane along straight lines with exactly
the same universal slope ($\Ms\propto \rs^{1.65}$) as in our
experiment and ended up at $z=0$ lying along another straight line,
with a different slope ($\Ms\propto \rs^{2.48}$) also identical to
that found in our experiment (see their Fig.~22). The fact that
  our constrained model reproduces the observed trend convincingly
  demonstrates haloes primarily grow inside-out.

\subsection{Radius Encompassing a Given Mass}\label{virrad}

As we will show, by assuming that haloes grow inside-out via PA and
that the spherical total energy in spheres of fixed mass is conserved
in the absence of shell-crossing, we are able to infer the radius
enclosing a given mass in a halo from the spherical energy
distribution of its progenitor protohalo.

To do this, we will consider the virial relation (\ref{vir0l}) for a
perfectly uniform sphere with mass $M$, which implies ${\cal
  W}(M)=-3GM^2/(5R)$, with spherical total energy ${\cal E}(M)$ equal
to that of the system at turnaround ${\cal E}\col (M)$ and null
spherical surface term ${\cal S}(M)$,
\beq 
R(M)= -\frac{3}{10}\,\frac{GM^2}{{\cal E}\col(M)}\,.
\label{vir0}
\eeq
This equation is the often used estimate for the radius encompassing
mass $M$ in spherical systems with an unknown internal mass
distribution (Bryan and Norman 1998). In principle, $R(M)$ is not
believed to give an exact measure of that mass in the real virialised
object as this has non-uniform density profile, its spherical total
energy is equal to ${\cal E}(M)$ instead of ${\cal E}\col(M)$ and its
surface term is not null. Yet, as shown below, as a consequence of
{\it the inside-out growth of accreting objects formed by PA}, the
inaccuracies above exactly cancel and both radii turn out to fully
coincide.

To see this, we will deform the system since its shells reach
turnaround so as to construct a virialised toy object satisfying the
conditions leading to equation (\ref{vir0}). This is only possible in
PA where the equivalent radius of turnaround ellipsoids increases with
increasing time and the central virialised object grows
inside-out. Shells reaching turnaround can then be {\it virtually}
contracted one after the other without any crossing so as to match the
mass profile $M(r)$ (although not necessarily the ellipsoidal
isodensity contours) of the real virialised object developed until
some time $t$. By ``virtual'' motion we mean a motion of shells that
preserves their particle energy and angular momentum, but is
disconnected from the real timing of the system. Of course, such a
motion will not recover at the same time the axial ratios of the
isodensity contours of the virialised object, but we only need to
recover the mass profile, which by conveniently contracting each new
shell is guaranteed (there is one degree of freedom: the final
equivalent radius of the contracted ellipsoidal shell and one quantity
to match: the mass within the new infinitesimally larger radius). By
construction, the new toy object so built has the same mass profile
$M(r)$ (although not the total energy profile $E(r)$ due to the
different ellipsoidal mass distribution) as the virialised object. But
the spherical total energy in such a toy object is equal to that at
turn-around, ${\cal E}\col[M(r)]$, as there has been no shell-crossing
and ${\cal E}$ is conserved for the appropriate choice of $s^2(r)$
(see Sec.~\ref{general}).

Of course, such a toy object is not in equilibrium. But the quantity
${\cal E}\col(R)-\widetilde {\cal W}(R)$, with $\widetilde {\cal W}(R)$
equal to the potential energy of a homogeneous sphere with mass $M$,
is positive (see eq.~[\ref{2Ecol0}] below). Therefore, we can
virtually expand each ellipsoidal isodensity contour of the toy
object, from the centre out to the edge avoiding shell-crossing, so as
to end up with a uniform density equal to the mean density
$\bar\rho(R)$ of the real object inside $R$ and still have an excess
of spherical kinetic energy. Thus, this can be re-distributed over the
sphere and the radial and tangential components of the spherical
velocity variance, $\widetilde s(r)$, can be locally exchanged so as
to satisfy the spherical virial relation (\ref{vir2l}) at every
radius. This is again possible to achieve because there are two
degrees of freedom, $\widetilde s(r)$ and $\widetilde s\rad(r)$, and
two conditions to fulfil, the spherical total kinetic energy in excess
and the spherical virial equation.  In this way, we have built a
steady homogeneous toy object with radius, mass and total energy
respectively equal to $R$, $M$ and ${\cal E}\col(M)$ and with null
value of $\widetilde s\rad(R)$ (there is no density outside the sphere
because shells having not reached turnaround at $t$ have not been
contracted). Hence, this uniform toy object satisfies the spherical
virial relation
\beq 
\frac{2{\cal E}\col(R)}{\widetilde {\cal W}(R)}=1\,,
\label{2Ecol0}
\eeq 
which, given the equality ${\widetilde {\cal W}(R)}=-3GM^2/(5R)$,
implies equation (\ref{vir0}). 

The exact equation (\ref{vir0}) allows one to determine the
spherically averaged density profile of the virialised object from the
properties of its seed. To do this we must simply take into account
that, during the initial expanding phase of shells, there is no
shell-crossing, so the spherical total energy in spheres with fixed
mass is conserved. We can then replace, in equation (\ref{vir0}),
${\cal E}\col(M)$ by its value in the protohalo, ${\cal E}\p(M)$, and
obtain, by inversion of $r(M)$, the mass profile $M(r)$ and, through
equation (\ref{rhot}), the spherically averaged density profile
$\srho(r)$ of the virialised object from the spherical total energy
distribution in the seed.

Furthermore, as the functions $\srho(r)$, $s^2(r)$ and $s^2\rad(r)$ do
not depend on the shape of the object, we can think in the spherically
symmetric case to infer the latter two functions from the former one.
In such objects, orbits are purely radial because they collapse and
virialise radially. We therefore have $s^2(r)=s^2\rad(r)$. Equations
(\ref{sig2}) and (\ref{sigr2}) then lead to a differential equation
for ${\cal E}(r)$ that can be readily integrated for the boundary
condition ${\cal E}=0$ at $r=0$, the result being
\beq 
{\cal E}(R)=-R\int_0^R \der r\left[4\pi\,\srho(r)\,GM(r)+\frac{{\cal W}(r)}{2r^2}\right]\,.
\label{diff}
\eeq
Once $\srho(r)$ is known, we can calculate ${\cal W}(r)$
(eq.~[\ref{potential}] and then ${\cal E}(R)$ (eq.~[\ref{diff}]), and
apply any of the two relations (\ref{sig2}) and (\ref{sigr2}) to infer
the profile $s^2(r)$. But the interest of this expression is only
formal because, as mentioned, the profile $s^2(r)$ is not observable,
not even in the spherically symmetric case where $s(r)=\sigma(r)$ and
$s\rad(r)=\sigma\rad(r)$. The reason for this is that objects formed
by PA are never spherically symmetric. The tidal field of surrounding
density fluctuations causes spherically symmetric seeds to undergo
ellipsoidal collapse \citep{Z70}. Even if such a tidal field is
artificially removed, the system suffers radial orbit instability,
also leading to triaxial virialised objects \citep{Hea99,McM06} .

The expression (\ref{diff}) is nonetheless useful to compute the
spherical total energy dissipation factor since the time of the
protohalo, ${\cal D}(M)\equiv {\cal E}(M)/{\cal E}\p(M)$, with ${\cal
  E}\p(M)$ given by equation (\ref{vir0}). This has the following
interesting implication. In objects formed by PA, there is always some
spherical energy loss through shell-crossing during virialisation, so
${\cal D}(M)$ is greater than one. Taking into account equations
(\ref{vir0l}) and (\ref{2Ecol0}), this dissipation factor can be
written in the form
\beq
{\cal D}(M)=\frac{{\cal W}(M)}{\widetilde {\cal W}(M)}\left[1-{\cal S}(M)\right]\,.
\label{D}
\eeq 
Thus, if ${\cal D}(M)$ is greater than one, the ratio
\beq 
\frac{{\cal W}(M)}{\widetilde {\cal W}(M)}=\frac{5}{6}\left[1+\int_0^{R(M)}
\frac{\der r}{R(M)}\;
\frac{r^4\,\bar\rho^2(r)}{R^4\,\bar\rho^2(R)}\right]\,,
\label{Ftau}
\eeq
must also be greater than one\footnote{As ${\cal W}(M)/{\widetilde
    {\cal W}(M)}$ is positive (see eq.~[\ref{Ftau}]), the factor
  $1-\cal{S}(M)$ must also be positive and less than one.}. By
differentiation and after a little algebra it can be seen that ratio
${\cal W}(M)/{\widetilde {\cal W}(M)}$ is greater than one provided
only $\bar\rho(r)$ is outward-decreasing. Consequently, we are led to
the conclusion that the spherically averaged mean inner density
profile of any collisionless dissipationless virialised object formed
hierarchically (through PA but also through major mergers; see
Sec.~\ref{discussion}) {\it is necessarily outward-decreasing}.

\section{Protohaloes}\label{peaks}

According to the previous results, to calculate the spherically
averaged density profile for a CDM halo grown by PA we only need the
spherical energy distribution, ${\cal E}\col(M)$, of its seed, namely
a peak in the initial density field filtered at the mass scale of the
halo. The cosmic time $\ti$ where the initial density field has to be
considered is arbitrary although small enough for the protohalo to be
in linear regime.

Such a spherical energy distribution is given, in the parametric form,
by the spherical total energy of the protohalo,
\beqa 
{\cal E}\p(R\p)=4\pi\!\int_0^{R\p}\!\der r\p\, r\p^2\,\lav \rho\p\rav(r\p)\nonumber~~~~~~~~~~~~~~~~~~~~~~\\ 
\times\left\{\frac{\left[H\ii r\p-v\p(r\p)\right]^2}{2}+\frac{\sigma^2\p(R\p)}{2}-\frac{GM(r\p)}{r\p} \right\},
\label{E1}
\eeqa
in centred spheres of radii $R\p$ encompassing the masses $M$,
\beq 
M(R\p)=4\pi\int_0^{R\p} \der r\p\, r\p^2\, \lav \rho\p\rav(r\p)\,.
\label{M1}
\eeq 
In equations (\ref{E1}) and (\ref{M1}), $H\ii$ is the Hubble constant
at the cosmic time $\ti$ of the protohalo, $\lav \rho\p\rav(r\p)$ is
its spherically averaged density profile, $v\p(r\p)$ is the peculiar
velocity, to 0th-order in the deviations from sphericity, due to the
gravitational pull by the mass excess within the radius $r\p$ and
$\sigma\p(R\p)$ is the uniform and isotropic background peculiar
velocity dispersion inside the sphere of radius $R\p$\footnote{The
  starting value of the spherical total energy can be computed taking
  $s^2\p\equiv \sigma\p^2$.}. 

The peculiar velocity $v\p(r\p)$ is (e.g. \citealt{Peebles}),
\beq
v\p(r\p)=\frac{2G\left[M(r\p)-4\pi r\p^3\rho\ii/3\right]}{3H\ii r\p^2}\,,
\label{vp}
\eeq
being $\rho\ii$ the mean cosmic density at $\ti$\footnote{In equation
  (\ref{vp}), we have taken into account that the cosmic virial factor
  $f(\Omega)\approx \Omega^{0.1}$ is at $\ti$ very approximately equal
  to one.}. Bringing this expression of $v\p(r\p)$ into equation
(\ref{E1}) and neglecting the second order term in the perturbation
with $v^2\p(r\p)$ (the term with $\sigma^2(R\p)$ may be greater than
this owing to the contribution of $\sigma^2\DM(t_i)$), we are led to
\beq 
{\cal E}\p(R\p)=\frac{5}{3}\,{\cal E}\p\HH(R\p)+M(R\p)\,\frac{\sigma^2\p(R\p)}{2}\,,
\label{E2}
\eeq
where ${\cal E}\p\HH(R\p)$ stands for the spherical total energy of
the protohalo in the case of pure Hubble flow (i.e. eq.~[\ref{E1}]
with null $v\p$ and $\sigma\p$). In principle, $\sigma^2\p(R\p)$ has
two contributions: one arising from the dark matter particle
velocities at decoupling, adiabatically evolved until the time $\ti$,
$\sigma^2\DM(\ti)$, and another one caused by random density
fluctuations. Following the exact prescription by \citet{MDL07} 
for this latter contribution, we arrive at
\beqa
\sigma\p^2(R\p)=\sigma^2\DM(\ti)+\Sigma^2\ii \int_0^{\infty}\frac{\der
k}{2\pi^2}\,P(k)\,\left[1-{\rm e}^{-k^2R\p^2}\right]^2~\nonumber\\
=\sigma^2\DM(\ti) +\!\Sigma^2\ii\left[\sigma^2\m(0)-2\sigma^2\m(R\p)+\sigma^2\m(\sqrt{2}R\p)\right]\!,
\label{sigp}
\eeqa
where $\Sigma\ii$ is defined as $\sqrt{2}\pi H a^2\ii$, being $a\ii$
the initial cosmic scale factor\footnote{At small $\ti$, this is very
  approximately equal to the cosmic growth factor.}, and
$\sigma_{-1}^2(\R)$ is the spectral moment of order $-1$ for the
power-spectrum, $P(k)$, with the $j$-order moment given by
\beq
\sigma_j^2(\R)=\int_{0}^\infty \frac{\der k}{2\pi^2} \,P(k)\,k^{2j+2}\,{\rm e}^{-k^2\R^2}\,.
\label{spec}
\eeq

But the velocity variance due to random density fluctuations,
$\sigma\p^2(R\p)-\sigma^2\DM(\ti)$ (see eq.~[\ref{sigp}]) turns out to
be several orders of magnitude less than $GM(R\p)/R\p$ (a factor $\sim
10^{-6}$ in the concordance model adopted here). Consequently, it can
be neglected in ${\cal E}\p(M)$ (eq.~[\ref{E1}]). Contrary to the
claims made by \citet{H02}, \citet{As04} and \citet{MDL07} that such a
velocity dispersion may affect the central density profile for CDM
haloes. The former two authors reached this conclusion by analysing
the effects of velocity dispersions that, unlike those used here, were
not consistently inferred from the random density fluctuations.
\citeauthor{MDL07} used the same derivations for the dispersion as
used here, however, they considered the fluctuations within the whole
protohalo, whereas we account for random density fluctuations within
each sphere of radius $R\p$. In including fluctuations from the whole
protohalo, Mikheeva et al.  included velocities induced by {\it
  external} density fluctuations with masses possibly larger than the
own mass of the sphere. This overestimates the velocity dispersion in
the sphere because external density fluctuations cause it a bulk
velocity, not a velocity dispersion. And the bulk velocity of a sphere
does not hamper its collapse, whereas the internal velocity dispersion
does, as found by \citeauthor{MDL07} for small enough $R\p$. The only
velocity dispersion that does not diminish with decreasing $R\p$ and,
hence, can really set a minimum halo mass is that intrinsic of dark
matter particles, $\sigma^2\DM(\ti)$. But this is negligible in CDM
cosmologies.

And what about the spherically averaged peak density profile,
$\lav\rho\p\rav(r\p)$? The BBKS profile calculated by \citet{BBKS}
gives the typical (average) spherically averaged density contrast
profile for peaks {\it filtered} with a Gaussian
window. Unfortunately, the convolution by a Gaussian window results in
some information loss so the BBKS profile cannot be used to infer the
desired profile. In addition, the {\it average} profile of purely
accreting haloes with $M$ at $t$ may not coincide with the profile
arising from the {\it average} profile of the corresponding
seeds. However, we can still find the typical (in the sense below)
spherically averaged density profile of protohaloes.

\begin{figure}
\centerline{\includegraphics[scale=0.45]{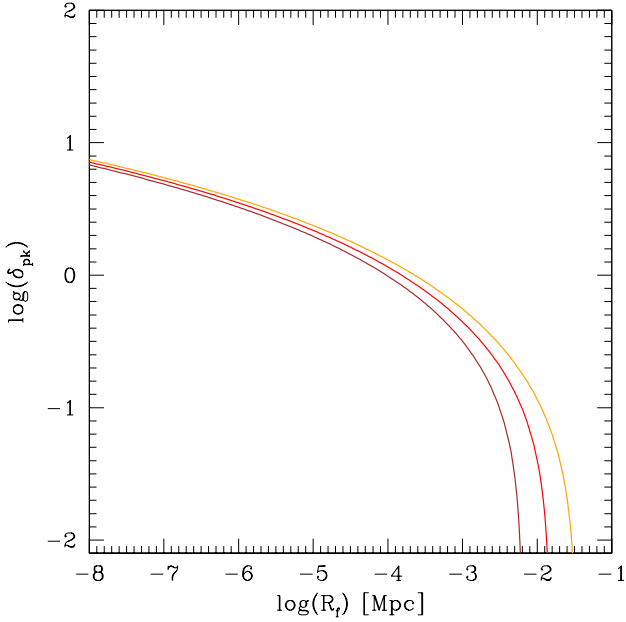}}
\caption{Typical $\delta\pk$--$\R$ (in physical units) peak trajectory
  of peaks at $z=100$ (solid lines) giving rise to the series of halo
  ancestors evolving by PA into haloes with current masses $M\tot$
  equal to 10, 1 and 0.1 times the critical mass for collapse,
  $M_\ast=3.6\times 10^{12}$ \modotc (curves from top to bottom,
    respectively in orange, red and brown) .}
\label{f1}
\end{figure}

As in PA haloes grow inside-out, the density profile inside every
inner ellipsoid exactly matches that of one halo ancestor, also
evolved by PA from a peak with its own density contrast and
scale. Hence, the typical (average) spherically averaged density
profile for purely accreting haloes with $M$ at $t$ must arise from a
protohalo whose spherically averaged density contrast profile, $\lav
\delta\p\rav (r\p)$, convolved with a Gaussian window of radius $\R$
corresponding to the mass of any halo ancestor yields, at $r\p=0$,
the density contrast $\delta\pk(\R)$ of the peak evolving into that
ancestor,
\beq 
\delta\pk(\R)=
\frac{2^{1/2}}{\pi^{1/2}\R^{-3}}\!\!\int_0^\infty\! \der
r\p r\p^2 \lav \delta\p\rav
(r\p)\exp\!\left[-\frac{1}{2}\left(\frac{r\p}{\R}\right)^{\!\!2}\right]\!\!.
\label{Fred}
\eeq 
According to the peak formalism (see Sec.~\ref{discussion}), the typical (most
probable) trajectory $\delta\pk(\R)$ of peaks evolving into the series
of halo ancestors ending in a halo with $M$ at $t$ is the solution of
the differential equation (SSb)
\beq
\frac{\der \delta\cs}{\der \R}= -x_{{\rm e}}(\delta\cs,\R)\,\sigma_2(\R)\,\R\,,
\label{dmd}
\eeq
with the boundary condition defined by the halo, that is satisfying
the relations
\beq 
\delta\pk(t)=\delta\co(t) \frac{G(\ti)}{G(t)}\,
\label{deltat}
\eeq
and
\beq
\R(M)=\frac{1}{q}\left[\frac{3M}{4\pi\rho\ii}\right]^{1/3}\,.
\label{rm}
\eeq
where $G(t)$ is the cosmic growth factor, $q=2.75$ is the radius, in
units of $\R$, of the collapsing cloud with volume equal to
$M/\rho\ii$, $\delta\co[t(z)]=1.93+(5.92-0.472 z+0.0546
z^2)/(1+0.000568 z^3)$ is the critical linearly extrapolated density
contrast for halo formation at the redshift $z$ (see Sec.~\ref{peaks}
for those values of $q$ and $\delta\co[t(z)]$). In equation
(\ref{dmd}), $x_{{\rm e}}(\delta\cs,\R)$ is the inverse of the average
inverse curvature $x$ (equal to minus the Laplacian over $\sigma_2$)
for the distribution of curvatures (BBKS),
\beq \left\lav
\frac{1}{x}\right\rav (\R,\delta\cs)\!=\! {(2\pi)^{-1/2} \over
  (1-\gamma^2)^{1/2}}\!\! \int_0^\infty\!\! \der x\,\frac{1}{x}\,
f(x)\,{\rm e}^{-{(x-x_\star)^2 \over 2(1-\gamma^2)}}\,,
\label{G}
\eeq
at peaks with $\delta\cs$ and $\R$, being
\beqa
f(x)=\frac{x^3-3x}{2}\left\{{\rm erf}\!\left[\left(\frac{5}{2}\right)^{1/2}x\right]+{\rm erf}\!\left[\left(\frac{5}{2}\right)^{1/2}\frac{x}{2}\right]\right\}\nonumber~~\\
+\left(\frac{2}{5\pi}\right)^{\!\!1/2}\!\!\left[\left(\!\frac{31x^2}{4}+\frac{8}{5}\!\right){\rm e}^{-\frac{5x^2}{8}}+\left(\!\frac{x^2}{2}-\frac{8}{5}\!\right){\rm e}^{-\frac{5x^2}{2}}\right]\!,
\label{fx}
\eeqa
and $\sigma_2(\R)$ the second order spectral moment, where $\gamma$
and $x_\star$ are respectively defined, in terms of the spectral
moments, as $\sigma_1^2/(\sigma_0\sigma_2)$ and $\gamma
\delta\cs/\sigma_0$. This distribution function is a very peaked,
quite symmetric, bell-shaped function so that the function
$\R(\delta\pk)$ inverse of the $\delta\pk(\R)$ solution of equation
(\ref{dmd}) is traced by peaks with the average (essentially equal to
the most probable) inverse curvature at each point
($\delta\pk,\R$). Note that the slope $\der \R/\der \delta\pk$
translates into the typical (average or most probable) accretion rate,
$\der M/\der t$, of haloes with $M(\R)$ at $t(\delta\pk)$ evolving
from those peaks. In Figure \ref{f1}, we show the typical peak
trajectories at $z=100$ leading to typical haloes with current masses
equal to $10^{-1}M_\ast$, $M_\ast$ and $10M_\ast$, where $M_\ast$ is
the critical mass for collapse in the concordance model ($3.6\times
10^{12}$ \modotc).

\begin{figure}
\centerline{\includegraphics[scale=0.45]{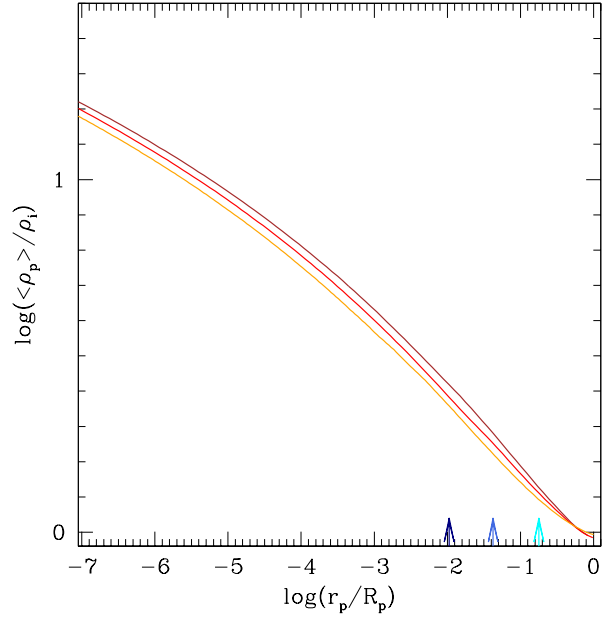}}
\caption{Spherically averaged unconvolved density profiles for the
  seeds of typical purely accreting haloes with the same current
  masses $M\tot$ as in Figure \ref{f1} (same colours, now ordered
    from bottom to top), obtained by inversion of the peak
  trajectories shown in that Figure. $R\p$ is the radius of the
  protoobject with mass $M\tot$. The filtering radii $\R/R_p$ used in
  the convolutions shown in Figure \ref{f3} are marked with arrows.}
\label{f2}
\end{figure}

Given a typical peak trajectory, $\delta\cs(\R)$, equation (\ref{Fred}) is a
Fredholm integral equation of first kind for $\lav\delta\p\rav
(r\p)$. Through the changes $y=r_p^2$ and $x=1/(2\R^2)$, it takes the
form of a two-sided Laplace integral transform,
\beq
g(x)=\int_{-\infty}^{\infty}\der y\,f(y)\,\mathrm{e}^{-xy}, 
\label{gx}
\eeq
with $f(y)$ and $g(x)$ respectively equal to $y^{1/2}\lav\delta\p\rav
(y^{1/2})$ and $2\sqrt{2\pi}(2x)^{-3/2}\delta\pk[(2x)^{-1/2}]$, which
can be solved in the standard way. Extending $x$ to the complex space
and taking $x=i2\pi\xi$, equation (\ref{gx}) adopts the form of a
Fourier Transform,
\beq
g(\xi)=\int_{-\infty}^{\infty}\der y\,f(y)\,\mathrm{e}^{-i2\pi\xi y},
\label{g}
\eeq
being
$g(\xi)=2\sqrt{2\pi}(i4\pi\xi)^{-3/2}[\delta_R(\xi)+i\delta_I(\xi)]$,
where $\delta_R(\xi)$ and $\delta_I(\xi)$ stand respectively for the
real and imaginary parts of $\delta\pk(\xi)\equiv
\delta\pk[(4\pi\xi)^{-1/2}]$. Thus, equation (\ref{Fred}) can be
inverted by simply taking the inverse Fourier transform of equation
(\ref{g}). 

\begin{figure}
\vskip -10pt
\centerline{\includegraphics[scale=0.45]{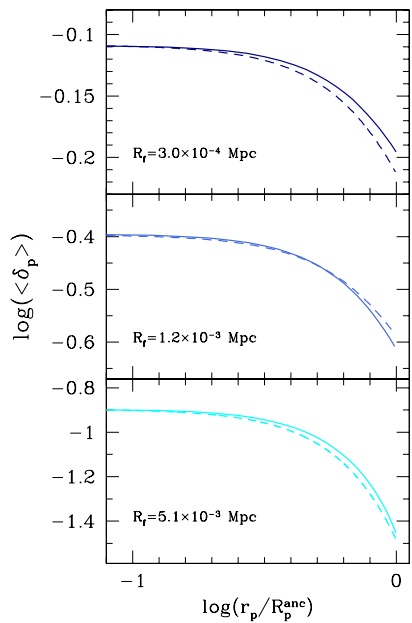}}
\caption{Spherically averaged density contrast profiles for the seed
  at $z=100$ of a halo with current mass equal to $M_\ast$ (in red in
  Figs.~\ref{f1} and \ref{f2}) convolved by a Gaussian window (solid
  lines) with the filtering radii quoted (in physical units) and
  marked with arrows in Figure \ref{f2} (same colours), compared to
  the BBKS profiles with identical central density contrasts and
  filtering scales (dashed coloured lines). $R\p^{{\rm anc}}$ are the
  radii of the seeds of the corresponding halo ancestors.}
\label{f3}
\end{figure}

As $f(y)$ is a real function, the real and imaginary parts of $g(\xi)$
must be even and odd, respectively. Both conditions must be satisfied
at the same time, so there are only two possibilities: either
$\delta_R(\xi)=\delta_I(\xi)$ or $\delta_R(\xi)=-\delta_I(\xi)$. The
latter possibility leads to a nonphysical solution, so $g(\xi)$ must
be a pure imaginary odd function,
i.e. $\delta_R(\xi)=\delta_I(\xi)$. To determine $\delta\pk(\xi)$ we
must solve the differential equation (\ref{dmd}) in the complex space
for $\xi$, that is for its real and imaginary parts separately. But,
as the two solutions $\delta_R$ and $\delta_I$ are identical, we can
directly solve it in the real space for $\R$ and then take
$\delta_R=\delta_I=\delta\pk/\sqrt{2}$, after the conversion from $\R$
to $\xi$. Once the solution $\delta\pk(\xi)$ is known, we can readily
calculate the function $g(\xi)$ and take its inverse Fourier
transform, which leads to the wanted protohalo density contrast
profile, $\lav\delta\p\rav (r\p)$, after the change $r_p^2=y$.

The spherically averaged unconvolved protohalo density profiles,
$\lav\rho\p\rav(r\p)$, obtained from the typical peak trajectories
depicted in Figure \ref{f1} are shown in Figure \ref{f2}. In Figure
\ref{f3}, we plot the convolved density contrast profiles for the
seeds of three arbitrary ancestors (with masses $M(\R)$ given by
eq.~[\ref{rm}] for the three arbitrary filtering radii) of the halo
with final mass $M_\ast$, corresponding to the filtering radii in
units of the protohalo radius, $\R/R_p$, marked with arrows in Figure
\ref{f2}. For comparison, we also plot the BBKS profiles for peaks
with identical central density contrasts and filtering radii. As can
be seen, each couple of curves is very similar; the small deviation
observed (of less than 10 \% within the radius $R\p^{\rm anc}$ of each
ancestor seed) is likely due to round-off errors\footnote{The
  convolution is achieved by taking the product of the 3-D Fourier
  transform of the unconvolved density contrast profile and the
  Gaussian window. As the Fourier transform of the unconvolved density
  contrast profile has no compact support, some aliasing is
  present.}. Thus, there is no evidence that the peak leading to an
object with the average spherically averaged density profile for
haloes with $M$ at $t$ is significantly different from the average
peak leading to such haloes.

Interestingly, the peak trajectories converge to a finite value (null
asymptotic slope) as $\R$ approaches to zero, as can be seen after
some algebra from the null asymptotic limit of the right hand-side
member of equation (\ref{Fred}). This in turn implies, from equation
(\ref{Fred}) and the relation between $\lav\delta\rho\p\rav(r\p)$ and
$\lav\rho\p\rav(r\p)$, that the unconvolved density profiles for
protohaloes with different masses also converge to a finite
value. This is consistent with the behaviour at small radii of the
unconvolved density profiles of protohaloes shown in Figure
\ref{f2}. This finite limit is approached slightly more rapidly than
in the case of typical peak trajectories.

\section{Halo Density Profile}\label{haloes}

Thus, to derive the typical spherically averaged density profile of a
halo with $M$ at $t$ we must follow the following steps: 1) solve
equation (\ref{dmd}) for the typical peak trajectory $\delta\pk(\R)$
leading by PA to the typical halo with $M$ at $t$; 2) from such a peak
trajectory, invert equation (\ref{Fred}) to find
$\lav\delta\p\rav(r\p)$ and from it the spherically averaged density
profile of the protohalo, $\lav\rho\p\rav(r\p)$; 3) given the
protohalo density profile, determine the spherical energy distribution
${\cal E}\p(M)$ in the protohalo by means of equations (\ref{M1}) and
(\ref{E1}); and 4) making use of this latter function, invert equation
(\ref{vir0}) to find the typical mass profile $M(r)$ of the halo and,
from it (eq.~[\ref{rhot}]), the typical spherically averaged density
profile $\srho(r)$.

\begin{figure}
\vskip 5pt
\centerline{\includegraphics[scale=0.45]{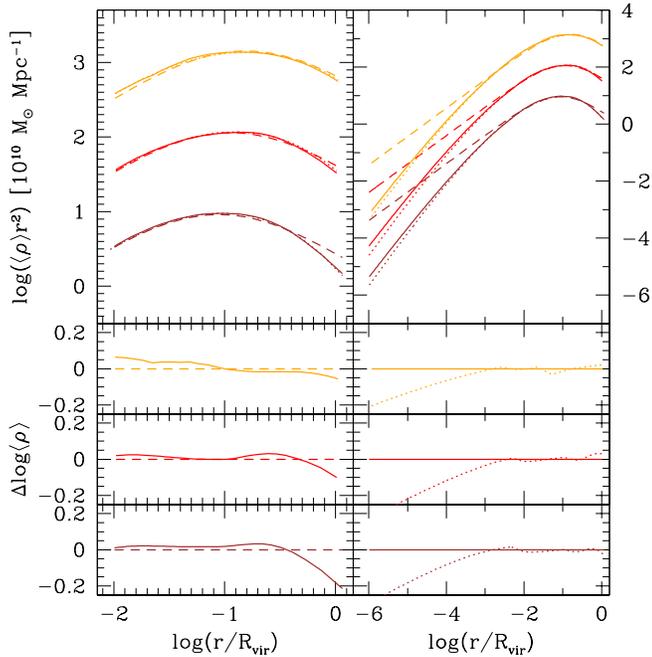}}
\caption{Spherically averaged density profiles predicted by the
  present model (solid lines) for the same current masses $M$ at
  $R\vir$ as in Figures \ref{f1} and \ref{f2} (same colours, 
    ordered as in Fig.~\ref{f1}), compared to the density profiles of
  the NFW form (dashed lines) with \citet{ZJMB09} concentrations
  fitting the average density profile of simulated haloes, both in the
  radial range covered by those authors (left) and down to a radius
  four orders of magnitude less (right). For comparison we also plot
  the fits to the predicted density profiles by an Einasto law (dotted
  lines). To avoid crowding, the curves for 10 $M_\ast$ and 0.1
  $M_\ast$ have been respectively shifted upwards and downwards by a
  factor of 3. The lower panels show the residuals of the theoretical
  curves from the NFW profiles (left) and of the Einasto profiles from
  the theoretical curves (right) for the three halo masses (same order
  from top to bottom).}
\label{f4}
\end{figure}

The density profiles so obtained (hereafter the theoretical profiles)
for haloes with the same masses as in Figures \ref{f1}--\ref{f2} are
compared in Figure \ref{f4} to the corresponding NFW profiles with
\citet{ZJMB09} mass-concentration and a total halo radius equal to the
virial radius $R\vir$ of \citet{bn98} (essentially equal to
$r_{90}$). As can be seen, there is good agreement between each couple
of curves down to about one hundredth $R\vir$. The residuals $\Delta
\log (\srho)$ (prediction minus NFW profile) start being positive at
$0.01R\vir$, tend to diminish at intermediate radii (reaching slightly
negative values for $M_\ast$), then increase again at moderately large
radii (except for 10 $M_\ast$) and finally become negative near the
halo edge. All these trends coincide with those shown by the same
residuals (simulated halo minus NFW profile) for individual haloes in
the simulations by \citeauthor{Navea04}~(\citeyear{Navea04}; Fig.~1,
left panels) except for the fact that the residuals of simulated
haloes are notably larger, as expected (the theoretical profiles are
supposed to correspond to average profiles). The only significant
difference is near the halo edge, where the residuals of the
theoretical profiles for low halo masses keep on decreasing for a
longer radial range (see also Fig.~\ref{f5}). This could be due
to the different halo radius used in both works: the radius adopted in
the present version of the model, arising from the fit to the halo
mass mass function predicted in the excursion set formalism (see
Sec.~\ref{discussion}) is likely closer to $R\vir\approx r_{90}$ as in
\citet{ZJMB09} than to $r_{200}$ as in \citet{Navea04}.

But, at small $r$, the theoretical profiles become progressively
shallower and increasingly deviate from the NFW form (with constant
inner logarithmic slope equal to $-1$). This behaviour is also in
agreement with the results of numerical simulations showing that the
Einasto profile gives a slightly better fit to the spherically
averaged density profile for simulated haloes down to radii of about
$0.001R\vir$. In Figure \ref{f4}, we plot the Einasto profiles that
better fit the theoretical profiles down to $0.01R\vir$. As can be
seen, these Einasto profiles are still reasonably close to the
theoretical profiles at radii four orders of magnitude less ($r\sim
10^{-6}R\vir$). 

\begin{figure}
\vskip 5pt
\centerline{\includegraphics[scale=0.45]{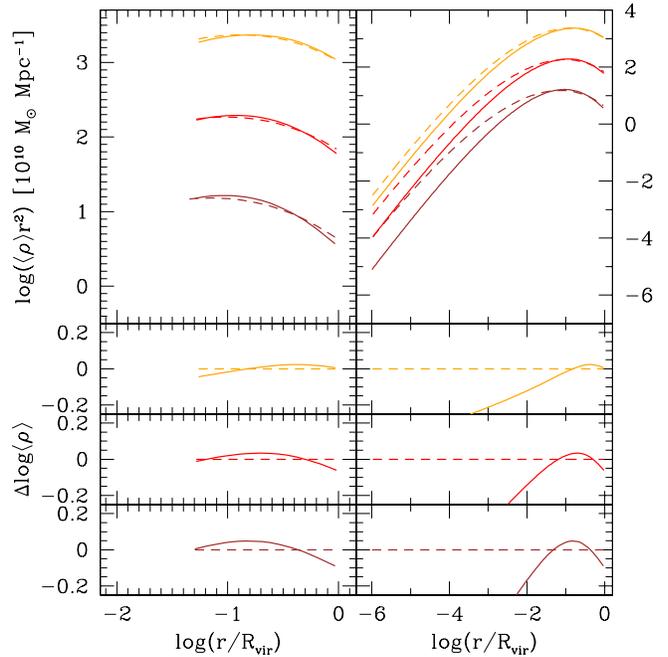}}
\caption{Same as Figure \ref{f4} (including colours and order)
  but for the comparison with the density profiles of the Einasto form
  (dashed lines) with \citet{Gao08} mass-dependent parameters fitting
  the average density profile of simulated haloes in the $\Lambda$CDM
  cosmology adopted by those authors, both in the radial range covered
  by \citeauthor{Gao08} although extended out to $R\vir\approx r_{90}$
  (left) and down to the same small radii as in the right panel of
  Figure \ref{f4} (right). The curves for 10 $M_\ast$ and 0.1
    $M_\ast$ have been shifted as in Figure \ref{f4}. The lower
  panels show the residuals (same lines) for the three halo masses
  (same order from top to bottom).}
\label{new}
\end{figure}

\begin{figure}
\vskip -28pt
\centerline{\includegraphics[scale=0.50]{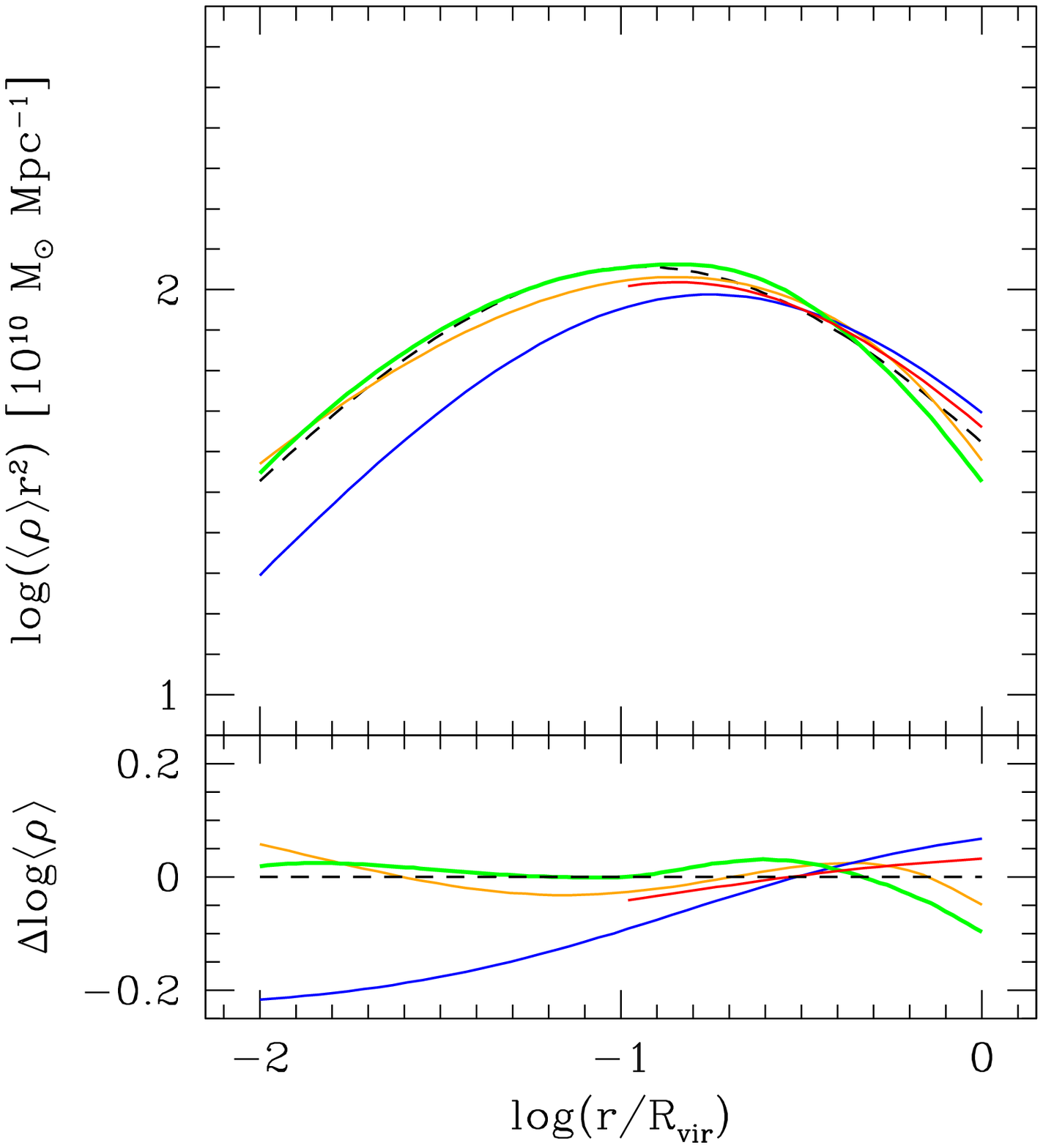}}
\caption{Spherically averaged density profile for a halo with current
  total mass inside $R\vir$ equal to $M_\ast$ predicted by the present
  model (green line), \citet{smgh07} model (orange line), the
    \citet{As04} model (red line) and the \citet{DPea00} model (blue
    line), respectively ordered from bottom to top at the right end of
    each panel. The \citeauthor{As04}'s profile is only shown at
  radii larger than 0.1$R\vir$ as recovered by those
  authors.}\label{f5}
\end{figure}

In Figure \ref{new}, the theoretical density profiles are also
compared to the Einasto profiles with mass-dependent parameters
according to \citet{Gao08} and the same cosmology used by these
authors (the mass-dependent Einasto parameters are not known for other
cosmologies). Although \citeauthor{Gao08} fitted the density profile
of simulated haloes only down to $0.05$ the total
radius\footnote{\citet{Gao08} used $r_{200}$. To transform to the same
  radius $R\vir$ as in Figure \ref{f4}, their Einasto profiles have
  been extended to $r_{90}$ where we have computed the halo mass.},
the agreement between each couple of curves is similarly good as for
the NFW profiles. In this case, the agreement of the theoretical
profiles with the Einasto profiles at very small radii is not so good
(although still much better than with the NFW profiles). But this is
likely due to the rather limited radial range covered in
\citeauthor{Gao08} study. As the NFW profile gives, down to $\sim
0.01R\vir$, similarly good fits as the Einasto profile to the density
profile of simulated haloes, the results shown in Figure \ref{f4}
strongly suggest that the Einasto profile can do much better at small
radii if the Einasto parameters are adjusted by covering a wider
radial range.

Given a protohalo with inner asymptotic density profile $\rho\propto
r\p^\alpha$, so that $M(r\p) \propto r\p^{(3+\alpha)}$ and ${\cal
  E}\p(r\p)\propto r\p^{(5+\alpha)}$ we have, from equation
(\ref{vir0}), that $\srho(r)\propto r^\alpha$. Since the protohalo has
a null logarithmic slope $\alpha$ (see Sec.~\ref{peaks}), it follows
that the density profile for haloes must also have null inner
logarithmic slope. In other words, there is strictly no cusp in the
spherically averaged density profile for CDM haloes according to the
present model. However, the finite central value is approached very
slowly, in fact slightly more slowly than the Einasto profile (see
Fig.~\ref{f4}).

The main differences the density profiles predicted by the present
model and by \citet{DPea00} and \citet{As04} models, both of which
also use the SI framework, are that we do not assume spherical seeds
with convolved BBKS profiles and we make use of inside-out growth
instead of an adiabatic invariant to determine $\srho(r)$. The model
presented here is essentially equivalent to the \citet{smgh07} model.
The key difference is that \citet{smgh07} used the typical
cosmological accretion rate onto haloes, whereas here we explicitly
make use of the typical density profile of halo seeds. However, the
typical halo mass accretion rate arises, as mentioned, from the
typical protohalo density profile, so this difference is just a matter
of presentation, motivated by the distinct theoretical framework of
the two models: the peak and the excursion set formalisms. The only
formal difference between the two models, apart from the fact that
\citeauthor{smgh07} also assumed spherical symmetry, is that the
radius encompassing a given mass adopted by these authors was inferred
from equation (\ref{vir0}) although not using the spherical total
energy of the real protoobject but of its {\it top-hat approximation}
according to \citet{bn98} prescription. This should introduce
additional numerical differences between the density profiles
predicted by the two models. It is also important to mention that the
models by \citet{DPea00} and \citet{As04} and \citet{smgh07} include
free parameters to be adjusted (the density contrast and filtering
radius of the peak in the two former cases, and the value of $\Delta
M/M$ setting the frontier between minor and major mergers in the
latter), while the present model includes no free parameter at all.

In Figure \ref{f5}, the spherically averaged halo density profile
predicted by the present model is compared to those predicted by the
\citet{DPea00}, \citet{As04} and \citet{smgh07} models, all of them in
the concordance model here considered. As expected, the theoretical
profile predicted by the present model is quite similar to that
predicted by \citeauthor{smgh07} model, while these two profiles
substantially deviate from those predicted by the remaining models. On
the other hand, the two former profiles are in better agreement with
the NFW profile with \citet{ZJMB09} mass-concentration relation.
Moreover, the theoretical profile derived here is the one in best
agreement with such a NFW profile, which is remarkable because it is
also the only theoretical profile with no free parameter to be
adjusted.

\section{IMPLICATIONS FOR MAJOR MERGERS}\label{discussion}

The present model has been built for haloes formed by PA. Not only is
this kind of halo formation crucial for the inside-out growth
condition, but also for the possibility to apply the peak formalism in
order to derive the peak trajectory leading to a typical halo with a
given mass at a given time. The peak formalism itself is based on the
existence of a one-to-one correspondence between haloes and peaks
inspired in the SI model that ignores major mergers. This therefore
raises two important questions. How do major mergers affect the
typical spherically averaged density profile for haloes derived under
these conditions? And how do they affect the peak formalism?

As shown in Section \ref{model}, given a seed with known spherically
averaged density profile, we can find the spherically averaged density
profile of the virialised halo evolving from it by PA, but the
converse is also true. Given a halo grown by PA, we can calculate from
its mass profile $M(r)$ the spherical total energy of the protohalo,
${\cal E}\p(M)$ (eq.~[\ref{vir0}]), and then determine its spherically
averaged density profile, $\lav\rho\p\rav(r)$, from equations
(\ref{E1}) and (\ref{M1}). Therefore, there is in PA {\it a one-to-one
  correspondence between the initial and final spherically averaged
  density profiles}. This is a well-known characteristic of {\it
  spherical} SI \citep{DPea00}, extended in the present paper to
non-spherical SI.

Interestingly, the reconstruction of the spherically averaged density
profile for the seed of a halo having grown by PA can also be applied
to a halo having suffered major mergers. This yields the spherically
averaged density profile, $\lav\rho\p\rav(r\p)$, of a putative peak
that would evolve by PA into a halo with a spherically averaged
density profile identical, by construction, to that of the original
halo. Clearly, if the halo has grown by PA, such a putative peak
exists and it is an ordinary peak. But if the halo has undergone major
mergers, does it exist? Is it an ordinary peak? In which halo does it
evolve? To answer these questions we will make use of the rigorous
treatment of the peak formalism given in MSSa and MSSb.

As mentioned, the peak Ansatz at the base of the peak formalism states
that there is a one-to-one correspondence between haloes with $M$ at
$t$ and peaks in the filtered density field at some small enough
cosmic time $\ti$, for some monotonous decreasing and increasing
functions $\delta(t)$ at $\R(M)$ of the respective
arguments. According to this Ansatz, peaks associated with {\it
  accreting} haloes describe continuous trajectories in the
$\delta\pk$--$\R$ diagram. Those peaks need not necessarily be
anchored to points with fixed coordinates; they can move as the
filtering scale varies (as real haloes do in the clustering
process). But, thanks to the {\it mandatory} use of the Gaussian
window\footnote{The decreasing behaviour of $\delta$ with increasing
  $\R$ implied by the growth in time of halo masses is only guaranteed
  for that particular filtering window (see MSSa).}, the connection
can be made in a simple consistent way between peaks tracing one given
accreting halo at contiguous scales\footnote{A peak at scale $\R$ is
  connected with another peak at scale $\R+\der \R$ provided only
  these two peaks are at a distance smaller than $\R$ from each other
  (see MSSa).}.

In a major merger, the continuous trajectories of (connected) peaks
tracing the merging haloes are interrupted, while one new continuous
peak trajectory appears tracing the halo resulting from the merger,
leaving a finite gap in $\R$ due to the skip in halo mass between the
merging and final objects. This is the only process where peak
trajectories are interrupted. Haloes that do not merge but are
accreted by more massive haloes are traced by peaks that do not
disappear but become nested into the collapsing cloud of larger scale
peaks tracing the accreting haloes. This leads to a complex nesting of
peaks with identical $\delta\pk$ but different $\R$. Once such a
nesting is corrected, the number density of peaks with $\delta\pk$ at
scales between $\R$ and $\R+\der\R$ and its filtering evolution
recovers the mass function (and growth rates) of virialised haloes
(MSSa and MSSb).

Thanks to these results it can be shown (see MSSa) that the
$\delta\pk(t)$ and $\R(M)$ relations defining the one-to-one
correspondence between (non-nested) peaks and (non-nested) haloes
stated in the peak Ansatz are necessarily of the form
(\ref{deltat})--(\ref{rm}). These relations can be seen as the
generalisation of those of the same form found in top-hat spherical
collapse, with $\delta\co(t)$ and $q$ respectively equal to 1.686 and
1. This does not mean, of course, that these parameters must take the
same values in the peak formalism. On the contrary, the freedom in
$\delta\co(t)$ and $q$ makes it possible to account for the change in
the filtering window (Gaussian instead of top-hat) and possibly also
in the departure from spherical collapse in the real clustering
process\footnote{For this latter aspect to be true, we should adjust
  the mass function obtained form simulations or from the excursion
  set formalism with values of $\delta\co$ and $q$ better adjusting
  the former mass function. In the present paper we fit, however, the
  mass function predicted in the excursion set formalism with
  $\delta\co(t)=1.686$ and $q=1$, so the departure from non-spherical
  collapse cannot be accounted for.} For
$\delta\co[t(z)]=1.93+(5.92-0.472 z+0.0546 z^2)/(1+0.000568 z^3)$ and
$q=2.75$, the halo mass function predicted in the $\Lambda$CDM
concordance cosmology (after correction for nesting according to MSSa)
recovers the mass function derived from the excursion set formalism
from $z=0$ up to any arbitrarily large $z$\footnote{The mass function
  predicted by the excursion set formalism is not fully accurate,
  particularly at large $z$, so one should rather fit the mass
  function drawn from numerical simulations. But this is not important
  here.}. As the halo mass function at $t$ predicted in the peak
formalism is nothing but the filtering radius distribution for peaks
with $\delta\pk(t)$ (eq.~[\ref{deltat}]) transformed into the former
by means of the relation (\ref{rm}), this result implies that there is
indeed a one-to-one correspondence between peaks and haloes as stated
by the peak Ansatz.

The putative peak of a halo and its associated peak according to the
peak Ansatz have the same $\delta\pk(t)$ and $\R(M)$. Their filtering
evolution may be different: the trajectory in the $\delta\pk$--$\R$
diagram of the putative seed can always be traced down to any
arbitrarily small filtering radius, whereas the trajectory of the
associated peak can only be traced until reaching the filtering radius
corresponding to the last major merger. However, the values
$\delta\pk$ and $\R$ of a peak do not allow one to tell its `past'
filtering evolution because there is, in the filtering process, a kind
of `memory loss'. The Gaussian window, mandatory as mentioned for the
peak formalism, yields a strong correlation between very close scales,
which is welcome to carry out the connection between peaks at
contiguous scales. But, at the same time, it yields a loss of
correlation between scales different enough to encompass the gap
produced in major mergers. Thus, owing to this particular window,
peaks at a given scale do not know whether or not they have appeared
in some major merger (at a smaller scale). Therefore, the putative
seed of a halo coincides with its associated peak according to the
peak Ansatz and, as such, it is an ordinary peak. (It contributes to
the filtering scale distribution of peaks with a given $\delta\pk$,
regardless of the past history of the halo.)

Thus, according to the present model supported by the results shown in
Section \ref{haloes}, the spherically averaged density profile of a
halo having undergone major mergers must be indistinguishable from
that arising from the evolution by PA of an ordinary peak (the
putative peak) or, equivalently, there must be in the halo aggregation
process a memory loss similar to that mentioned above affecting the
filtering process of peaks. Such an implication has to do with the
fundamental debate on whether or not virialisation is a real
relaxation (e.g. \citealt{Henrik09}).

As noticed by \citet{DPea00}, the one-to-one mapping between the
initial and final density profiles in PA seems to indicate that there
is no memory loss, at least in PA, during virialisation. This is at
odds, however, with the fact that virialisation, even in PA, sets a
time arrow. Even though the equations of motion of individual
particles are time-reversible, owing to the highly non-linear dynamics
of shell-crossing, any infinitesimal inaccuracies in the positions and
velocities of particles in a simulation are rapidly amplified, causing
simulated orbits to deviate from the true ones. When the simulation
is run forwards, this goes unnoticed because the system always reaches
the same (statistically indistinguishable) final equilibrium
configuration. However, when the simulation is run backwards it is
readily detected: before shells can reach the maximum size (at
turnaround), they begin to contract again towards the final
equilibrium state. As the existence of a time arrow (or the
impossibility of reversing by numerical means the evolution of a
system) is an unambiguous signature of relaxation, we must conclude
that virialisation is a real relaxation. 

In the case of PA the memory loss on the initial conditions is not
complete: the order of shell apocentres is conserved. Consequently,
despite the fact that information regarding individual orbits is lost,
we can recover the initial configuration in a statistical sense (in
that we can recover the density profile). Thus, even in this case, the
initial configuration cannot be {\it exactly} recovered because of the
loss of information on the phase of particle orbits (individual orbits
are mixed up). And, in the general case including major mergers, the
relaxation nature of virialisation is even more evident. The initial
configuration of a given virialised object can never be recovered, not
even in a statistical sense, because the density profile for
virialised haloes does not harbour information on their past history,
so we can never be sure that the unconvolved density profile of their
putative seeds, derived assuming PA, describes the density profile of
their real seeds.

The prediction made by the present model that the spherically averaged
density profile of haloes does not harbour information on their past
aggregation history is thus consistent with the general behaviour of
virialisation. Moreover, it is consistent with several specific
results regarding the density profile of simulated virialised haloes:
i) virialised haloes with very different aggregation histories have
similar spherically averaged density profiles (e.g. \citealt{NS99}),
ii) their density profile is independent of the time they suffered
their last major merger \citep{Wea02} and iii) their density profile
does not allow one to tell how many, when and how intense the major
mergers they have suffered are \citep{RD06}. Certainly, there are also
some claims in the literature that the density profile for haloes
depends on their formation time. But this is due to the particular
definition of halo formation time adopted in those works, related to
the form of the profile. Thus, the prediction of the model regarding
the lack of information in the density profile of a halo on its past
history is not contradicted by the results of numerical simulations.

Thus, the present model strongly suggests that haloes having suffered
major mergers have density profiles which are indistinguishable from
those that have not. This prediction is important not only for the
validity of the model (i.e. the validity of the inside-out growth
assumption and the peak Ansatz), already supported by the good
behaviour of the density profiles it predicts (Sec.~\ref{haloes}), but
also because it allows one to understand why major mergers do not
affect neither the typical halo density profile nor the peak formalism
for halo statistics based on SI. Indeed, the fact that haloes formed
in a major merger have, after complete virialisation, would have
a density profile indistinguishable from that of a halo formed by PA
justifies that any halo with $M$ at $t$ can be associated with a peak
with $\delta\pk$ and $\R$ as if it had evolved from it by PA, as
assumed in the peak Ansatz and that the typical halo density
profile derived assuming PA is not affected by major mergers. As
a corollary we have that the present model of spherically averaged
density profile should hold for all haloes, regardless of their
aggregation history.

\section{SUMMARY}\label{summ}

The present model relies on two assumptions: i) that the typical
density profile for virialised objects in hierarchical cosmologies
with collisionless dissipationless dark matter can be derived as if
all the objects grew by PA and ii) that the peak formalism correctly
describes halo statistics. The key points of the derivation are that,
contrarily to the kinematic profiles, the spherically averaged density
profile does not depend on the (triaxial) shape of the object and
that, during accretion, virialised objects grow from the inside out,
keeping the inner structure essentially unchanged.

Triaxial collisionless dissipationless systems undergoing PA virialise
by transferring energy from inner to outer shells through
shell-crossing. Due to this energy loss, particles progressively
reaching turnaround describe orbits that contract orderly in the sense
that the ellipsoidal surfaces effectively traced by their apocentres
shrink and change their axial ratios without crossing each other until
they stabilise. This causes the central steady object to develop
inside-out. Such an evolution of accreting objects is in full agreement
with the results of numerical simulations.

One important consequence of the inside-out growth is that the radius
encompassing a given mass in triaxial virialised objects formed by PA
exactly coincides with the usual estimate of such a radius from the
energy of the sphere with identical mass at turnaround. In this
conditions, there is a one-to-one mapping between the density profile
of the virialised object and its seed. This allows to infer the
typical spherically averaged density profile, $\srho(r)$, of
virialised objects from the energy distribution of their spherically
averaged seeds, which can be calculated from the power-spectrum of
density perturbations making use of the peak formalism.

The consistency between the peak formalism and the one-to-one mapping
between the initial and final density profiles following from PA
implies that virialisation must be a real relaxation. This conclusion
agrees with the general behaviour of virialisation (it sets a time
arrow) as well as with specific results regarding the density profile
of simulated haloes. As such, the spherically averaged density profile
for haloes cannot harbour information on their past aggregation
history which explains why major mergers do not alter the typical
density profile of virialised objects derived under the PA assumption
and do not invalidate the peak Ansatz. In other words, the sole
condition that virialisation is a real relaxation is enough for the
two assumptions of the model to be not only consistent but also fully
justified.

The model has been applied to CDM haloes. We have derived the typical
unconvolved spherically averaged density profile for peaks evolving by
PA and from it the typical spherically averaged density profile for
haloes with a given mass in a given epoch. Specifically, we have
established the link between the typical halo density profile and the
power-spectrum of density perturbations in any given hierarchical
cosmology. The typical halo density profile so predicted in the
$\Lambda$CDM concordance cosmology is in very good agreement with the
NFW and the Einasto profiles fitting the spherically averaged density
profile of simulated haloes down to one hundredth the total
radius. However, such a theoretical profile does not have a central
cusp. In the $\Lambda$CDM cosmology, the model predicts that
simulations reaching increasingly higher resolutions will find a
typical halo density profile that tends to a core (null asymptotic
slope) which is slowly approached, even slower than current fits using
the Einasto profile.

\vspace{0.75cm} \par\noindent
{\bf ACKNOWLEDGEMENTS} \par

\noindent This work was supported by the Spanish DGES,
AYA2006-15492-C03-03 and AYA2009-12792-C03-01, and the Catalan DIUE,
2009SGR00217. JV was beneficiary of the grant BES-2007-14736 and SS of
a grant from the Institut d'Estudis Espacials de Catalunya. We
acknowledge Gary Mamon and Guillermo Gonz\'alez-Casado for revising
the original manuscript and an anonymous referee for his constructive
criticism.


\end{document}